\documentclass[prc,twocolumn,twoside,showpacs,floatfix]{revtex4-1}

\usepackage{graphicx,color,rotati	ng}
\usepackage{amsmath,amssymb,bm}
\usepackage{dcolumn}

\usepackage{times,txfonts}

\usepackage{multirow}

\usepackage{pifont}
\usepackage{dsfont}

\newcommand{\eq}[1]{\begin{equation}#1\end{equation}}
\newcommand{\eqmulti}[1]{\begin{equation}\begin{split}#1\end{split}\end{equation}}

\newcommand{\comm}[2]{\ensuremath{[{#1},{#2}]}}

\newcommand{\clebsch}[6]{ \ensuremath{\left(\!\!
\begin{array}{cc}
 {#1} & \!\!\!\!{#2} \\
 {#4} & \!\!\!\!{#5} 
\end{array}
 \!\!\right|
\left.\!\!\!
\begin{array}{c}
 {#3}\\
 {#6}
\end{array}\!\! \right)}
}

\newcommand{\elem}[2]{\ensuremath{{}^{#2}\text{#1}}}

\definecolor{FGViolet}{rgb}{0.61,0.32,0.61}
\definecolor{FGDarkBlue}{rgb}{0,0,0.6}
\definecolor{FGBlue}{rgb}{0,0,0.8}
\definecolor{FGLightBlue}{rgb}{0.2, 0.6, 0.8}
\definecolor{FGGreen}{rgb}{0.2,0.7,0.2}
\definecolor{FGLightGreen}{rgb}{0.4,1,0.4}
\definecolor{FGYellow}{rgb}{1,0.95,0}
\definecolor{FGOrange}{rgb}{0.95,0.5,0.1}
\definecolor{FGRed}{rgb}{0.8,0,0}
\definecolor{FGWhite}{rgb}{1,1,1}
\definecolor{FGLightGray}{rgb}{0.8,0.8,0.8}
\definecolor{FGGray}{rgb}{0.5,0.5,0.5}
\definecolor{FGDarkGray}{rgb}{0.3,0.3,0.3}
\definecolor{FGBlack}{rgb}{0,0,0}

\newcommand{\LamChi}{\ensuremath{\Lambda_{\chi}}}
\newcommand{\LamSFR}{\ensuremath{\tilde{\Lambda}_{\chi}}}
\newcommand{\LamSRG}{\ensuremath{\lambda}_{\text{SRG}}}
\newcommand{\opt}{\ensuremath{\text{N}^2\text{LO}_{\text{opt}}}}

\begin{document}

\title{Sensitivities and correlations of nuclear structure observables emerging\\ from chiral interactions}

\author{Angelo Calci}
\affiliation{TRIUMF, 4004 Wesbrook Mall, Vancouver, British Columbia, V6T 2A3, Canada}
\email{calci@triumf.ca}

\author{Robert Roth}
\affiliation{Institut f\"ur Kernphysik, Technische Universit\"at Darmstadt,
64289 Darmstadt, Germany}
\email{robert.roth@physik.tu-darmstadt.de}

\date{\today}

\begin{abstract}  
Starting from a set of different two- and three-nucleon interactions from chiral effective field theory, we use the importance-truncated no-core shell model for \emph{ab initio} calculations of excitation energies as well as electric quadrupole (E2) and magnetic dipole (M1) moments and transition strengths for selected p-shell nuclei. We explore the sensitivity of the excitation energies to the chiral interactions as a first step towards and systematic uncertainty propagation from chiral inputs to nuclear structure observables. The uncertainty band spanned by the different chiral interactions is typically in agreement with experimental excitation energies, but we also identify observables with notable discrepancies beyond the theoretical uncertainty that reveal insufficiencies in the chiral interactions.  
For electromagnetic observables we identify correlations among pairs of E2 or M1 observables based on the \emph{ab initio} calculations for the different interactions. We find extremely robust correlations for E2 observables and illustrate how these correlations can be used to predict one observable based on an experimental datum for the second observable. In this way we circumvent convergence issues and arrive at far more accurate results than any direct \emph{ab initio} calculation. A prime example for this approach is the quadrupole moment of the first $2^+$ state in \elem{C}{12}, which is predicted with an drastically improved accuracy.   

\end{abstract}


\maketitle

\section{Introduction}
Over the past decade there has been substantial progress in the construction of nuclear forces from chiral effective field theory (EFT), both, on the formal level and on practical aspects~\cite{Epel06,EpHa09,MaEn11}. 
Recently several different regularization schemes have been implemented and are being explored in many-body calculations. An example are coordinate-space regulators leading to fully local two-nucleon (NN) and three-nucleon (3N) interactions up to N$^2$LO that can be used in quantum Monte Carlo calculations~\cite{GeTe13}. Using a mixed local and nonlocal regularization scheme Epelbaum {\it et al}. have presented a new family of improved chiral NN interactions ranging from leading-order (LO) to next-to-next-to-next-to-next-to leading order (N$^4$LO) with five different cutoff values \cite{EpKr15,EpKr15b}. This family of interactions allows for a systematic study of order-by-order convergence and cutoff dependence of nuclear structure observables, a critical aspect that was often ignored in previous nuclear structure applications. The LENPIC collaboration~\cite{LENPIC} is exploring these interactions in few- and many-body calculations \cite{BiCa15} and is  developing the consistent chiral 3N interactions. In a complementary development, new fitting strategies for chiral NN+3N interactions at N$^2$LO are exploited to quantify the statistical uncertainties related to the parameter fits~\cite{CaEk16}.  Moreover, novel fit procedures are utilized that improve the description of bound-state properties for nuclei beyond the few-body domain~\cite{EkBa13,EkJa15} compared to the previous generation of chiral interactions~\cite{EnMa03,Navr07,EpGl04,EpGl05,EpNo02,BeEp08}. There are also efforts to include the $\Delta$ resonance as an explicit degree of freedom to accelerate the convergence of the  chiral order expansion~\cite{KrEp07}.
 
These developments on chiral interactions enable numerous applications in nuclear structure physics. 
To probe the predictive power of chiral interactions without introducing uncontrolled approximations, \emph{ab initio} many-body approaches are the methods of choice. In addition to the traditional \emph{ab initio} many-body methods such as the no-core shell model (NCSM)~\cite{BaNa13,NaQu09,NaGu07} and the Green's function Monte Carlo (GFMC) method~\cite{PiWi01,WiPi02,PiWi04} there are also recent developments like the coupled-cluster (CC) methods~\cite{TaBa08a,ShBa09,HaPa10,HaHj12,BiPi13,BiLa13,BiLa13b}, the self-consistent Green's function methods~\cite{SoDu11,CiBa13,SoCi14}, and the in-medium similarity renormalization group (IM-SRG)~\cite{TsBo11,HeBo13,HeBi13} that extend the range of \emph{ab initio} nuclear structure calculations to medium-mass and heavy nuclei regime up to the tin isotopes. 
The importance truncated no-core shell model (IT-NCSM)~\cite{Roth09,RoNa07} bridges the gap between the traditional and novel many-body methods, it can include the 3N interaction explicitly and can probe ground-state and excitation energies as well as spectroscopic observables in p- and lower sd-shell nuclei. These observables constitute a comprehensive testbed for the theoretical predictions of chiral EFT.

Typically, these \emph{ab initio} approaches attempt to estimate uncertainties resulting from truncations and incomplete convergence with respect to the many-body space. However, in many applications of nuclear structure theory, such as the p-shell spectroscopy, the uncertainties entering through the chiral inputs have not been explored, so far. The combination of reliable many-body approaches and new chiral interactions will allow for a systematic propagation of theory uncertainties to the nuclear structure observables. As a preparatory step, we study sensitivity and correlations of different spectroscopic observables for a set of chiral NN+3N interactions. We explore the excitation spectra of \elem{Li}{6}, \elem{B}{10}, and \elem{C}{12} and quantify the sensitivity of excitation energies on the choice of the underlying chiral interactions. The variation of the underlying interactions also provides an opportunity to detect and map-out correlations among pairs of nuclear structure observables, particularly electromagnetic moments and transition strengths.

This paper is structured as follows. In Sec. \ref{sec:ham} we introduce the different chiral interactions and the technical aspects of the many-body treatment. The sensitivity analysis of the excitation spectra for selected p-shell nuclei is presented in Sec.~\ref{sec:spectra}. 
In Sec.~\ref{sec:elmagObservables} correlations in electromagnetic observables are studied and we conclude in Sec.~\ref{sec:conclusion}.

\section{From chiral Hamiltonians to observables }
\label{sec:ham}
\subsection{NN+3N interactions from chiral EFT}
\label{subsec:EFT}

In this work we investigate interactions from three different chiral schemes that are obtained at N$^2$LO or N$^3$LO using different regularization and fit procedures. 
The first NN interaction we introduce is developed by Entem and Machleidt (EM)~\cite{EnMa03} at N$^3$LO. The nonlocal regulator function uses a fixed cutoff of $500\,\text{MeV}/\text{c}$. This potential provides an accurate description of NN phase shifts with a comparable precision as more phenomenological high-precision potentials like Argonne V18~\cite{WiSt95} and CD-Bonn~\cite{Mach01}. The EM potential is widely used in nuclear structure physics and has been a standard choice in the \emph{ab initio} field.

The  $\opt$ interaction by Ekstr\"om {\it et al}.~\cite{EkBa13} is a recently developed chiral interaction at N$^2$LO. The fits of the low-energy constants (LECs) have been performed with the practical optimization using no derivatives (for squares) (POUNDerS) algorithm~\cite{KoLe10}. This potential also uses a cutoff of $\LamChi=500\,\text{MeV}/\text{c}$ and an additional spectral function regularization (SFR) with a cutoff of $\LamSFR=700\,\text{MeV}/\text{c}$.

The NN interaction by Epelbaum, Gl\"ockle, Mei\ss ner (EGM)~\cite{EpGl04} at  N$^2$LO uses the same non-local regularization with an additional SFR cutoff.
This potential is constructed for a sequence of five cutoff combinations $(\LamChi / \LamSFR)=\{ ( 450/500),( 600/500),( 550/600),( 450/700),$ $(600/700) \}  \,\text{MeV}/\text{c}$ and provides a slightly less precise reproduction of the NN data than the other two NN interactions.
The pion-nucleon LECs $c_i$ are fitted independently of the regularization to the pion-nucleon scattering data~\cite{BuMe00} and differ from the values used for the EM and $\opt$ interactions.
With the sequence of cutoff parameters it offers the unique opportunity to study the effect of the regularization on nuclear structure observables and, thus, to draw conclusions about the theoretical uncertainties originating from the chiral inputs. 

For chiral interactions with non-local regulators the cutoff variation provides a legitimate diagnostic tool to estimate the uncertainties at an individual chiral order. Nevertheless, a variation of the chiral order is crucial to study the convergence of the interactions and future works will combine a chiral order and cutoff variation for a more elaborate uncertainty analysis.
Note, for chiral interactions with local regulators, physical observables show generally a small sensitivity to variations in the cutoff~\cite{GeTe13,GeTe14,EpKr15}. Thus, novel studies with a semi-local regularization must include information of the chiral order convergence to extract the uncertainties of the currently available NN forces~\cite{BiCa15}. 

The above NN forces are augmented by 3N forces at N$^2$LO. 
The EM and $\opt$ NN forces are combined with the local 3N force using a cutoff of $500\,\text{MeV}/\text{c}$~\cite{Navr07}. 
The LECs $c_{1,3,4}$ of the two-pion exchange term are adopted from the NN interaction. The parameter $c_{\text{D}}$ is fitted to the triton $\beta$-decay half-life~\cite{GaQu09} and $c_{\text{E}}$ is fixed by the $A=3$ and \elem{He}{4} binding energy, for the EM and $\opt$ NN interaction, respectively. This yields $(c_{\text{D}},c_{\text{E}})=(-0.2 ,-0.205)$ for the EM interaction and $(-0.39, -0.398)$ for the $\opt$ interaction. 
Although, the $\opt$ interactions is originally a NN force for brevity we use this expression also to refer to the corresponding NN+3N interaction introduced in this work. 
The EGM NN forces at N$^2$LO are typically combined with a consistent non-local 3N force at N$^2$LO. While this NN+3N force is used in several applications to neutron matter~\cite{HeFu13,TeKr13,KrTe13}, nuclear structure physics beyond the lightest nuclei is fairly unknown. 
The LECs of the 3N force are fitted to the triton ground-state energy and the neutron-deuteron doublet scattering length~\cite{EpNo02}.  
The  partial-wave decomposed 3N matrix elements at N$^2$LO can be derived explicitly~\cite{Navr07,HuWi97,EpNo02} or via a numerical partial-wave decomposition~\cite{SkGo11,HeKr15}. The latter approach is also applicable to compute the complicated 3N contributions at N$^3$LO for future investigations. 

\subsection{SRG evolution and basis transformations}
\label{subsec:SRG}

Although non-local chiral NN interactions are rather soft due to the momentum cutoff in the regularization, it is still difficult to converge NCSM-type calculations beyond the lightest nuclei. Also the inclusion of the relevant 3N contributions can be problematic for \emph{ab initio} methods when a bare chiral interaction is used. The similarity renormalization group (SRG)~\cite{Wegn94,BoFu07,SzPe00} is a unitary transformation that softens the nuclear interaction and can be applied consistently in the two- and three-body space. Therefore, this approach is used in a variety of nuclear structure applications to soften the chiral NN+3N interactions~\cite{SoCi14,HeBo13,BiPi13,RoLa11,JuNa11,JuNa09}.

The SRG flow equation for the Hamiltonian $H$ is given by 
\eq{ \label{eq:srg_floweq}
  \frac{d}{d\alpha} H_{\alpha} = \comm{\eta_{\alpha}}{H_{\alpha}} \;,
}
with the continuous flow-parameter $\alpha$, which is related to a momentum scale $\LamSRG=\alpha^{-1/4}$ and the dynamic generator  
\eq{ \label{eq:srg_eta}
  \eta_{\alpha} = (2\mu)^2\; \comm{T_{\text{int}}}{H_{\alpha}}\;,
}
where $\mu$ is the reduced nucleon mass and $T_{\text{int}}$ is the intrinsic kinetic-energy operator.
It is important to note that the SRG evolution induces irreducible many-body contributions beyond the particle rank of the initial interaction.
With the canonical generator~\eqref{eq:srg_eta} it has been found~\cite{RoLa11,JuNa11,JuNa09}, that it is indispensable to include the induced three-body contributions. Therefore, for all results presented in this paper we use an initial NN and NN+3N interaction and include all contributions up to the three-body level, which correspond to the NN+3N-induced and NN+3N-full nomenclature of previous works~\cite{RoCa14,RoLa11,JuNa11}.   
      
We aim at many-body calculations performed in the harmonic-oscillator (HO) representation. Thus the NN and 3N interactions that are obtained in a partial-wave momentum representation need to be transformed to the HO space. There are techniques to perform the evolution equation of the SRG in the three-body momentum representation~\cite{Hebe12}, however, for applications in localized systems, such as nuclei, the discrete HO Jacobi basis is the most efficient scheme for the SRG evolution.
Therefore, we immediately transform the 3N momentum matrix elements to the HO Jacobi representation, performing the SRG transformation subsequently. 

The HO machinery, comprehensively described in~\cite{RoCa14}, is utilized to perform the Moshinsky transformation to the particle-basis representation. Eventually, the matrix elements are stored in the so-called $JT$-coupled scheme~\cite{RoCa14,RoLa11}, which is the starting point for a number of \emph{ab initio} many-body methods~\cite{LaNa15,BoHe14,SoCi14,WiGa14,CiBa13,HeBi13,HeBo13,BiLa13,RoBi12,RoLa11}.


\section{Excitation Spectra \label{sec:spectra}}

In a first step we study the excitation spectra of various p-shell nuclei using the different chiral Hamiltonians introduced in Sec.~\ref{subsec:EFT}.
The focus will be on the sensitivity of the different excited states on the chiral inputs, giving rise to systematic theory uncertainties that result from the various choices made during the construction of the chiral interactions. This includes, for instance, the different regularization schemes, chiral orders as well as the fit procedures used for the LECs. An additional source of uncertainty are the statistical uncertainties of the LECs resulting from the fits. The latter uncertainties have been exploited recently for few-body scattering and ground-state observables~\cite{CaEk16}, but remain to be investigated for p-shell spectroscopy.

We employ the IT-NCSM with the SRG evolved Hamiltonians based on the chiral NN or  NN+3N interactions discussed in Sec.~\ref{subsec:SRG}. For all Hamiltonians the IT-NCSM calculation includes explicit 3N terms of the SRG-evolved Hamiltonians. We perform a full NCSM calculations up to $N_{\max}=6$  for \elem{Li}{6} and $N_{\max}=4$ for \elem{C}{12} and \elem{B}{10}. We efficiently proceed to larger $N_{\max}$ with an importance truncation including a threshold extrapolation towards the full NCSM space. The details of the IT-NCSM and the threshold extrapolation are discussed in~\cite{Roth09,RoNa07}. On should note that the threshold extrapolation itself induces a theoretical uncertainty at the level of the final observables, which is quantified systematically through the extrapolation protocol~\cite{Roth09}. For excitation energies this uncertainty is of the order of $50\,\text{keV}$  for the largest spaces. The IT extrapolation is very robust for the discussed excitation spectra in this work. The only exception is the first excited $0^{+}$ state in \elem{C}{12} for a single EGM interaction, where the degeneracy with the first $4^{+}$ state causes inaccurate IT extrapolations, resulting in an unreliable assessment of the angular momentum of this state.

\begin{figure*}[t]

\hspace*{-20pt}\includegraphics[width=1.1\textwidth]{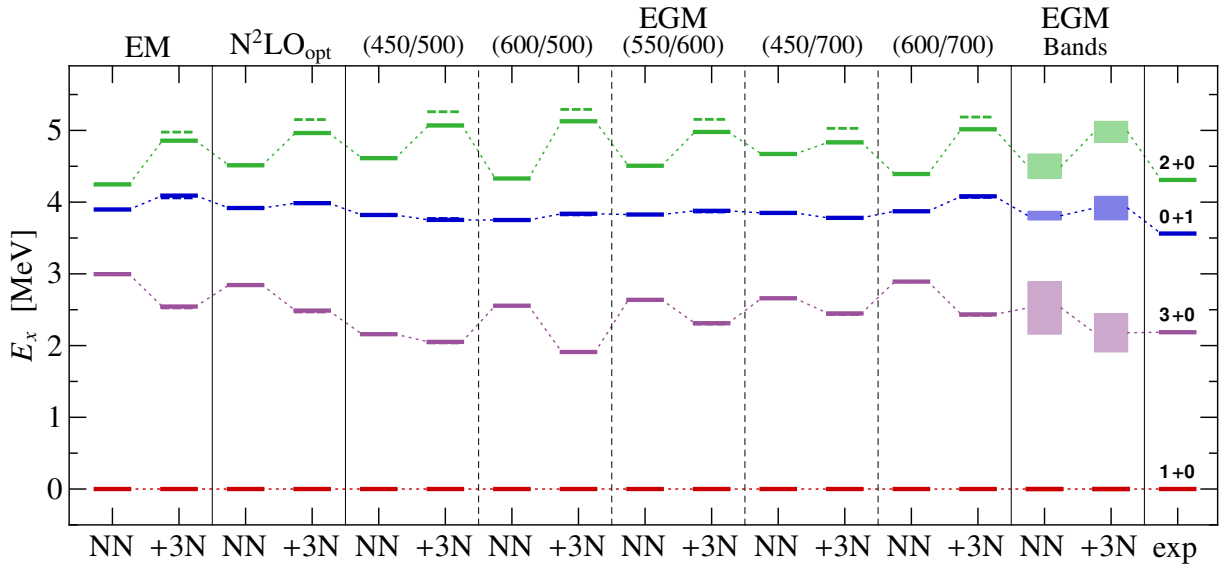}\\[-8pt] 
\caption{(color online) Excitation spectrum of \elem{Li}{6} for the EM, the $\opt$ and all five cutoff combinations of the EGM NN and NN+3N interactions. The parameters of the IT-NCSM calculations are $N_{\text{max}}=10$, $\hbar\Omega=16\,\text{MeV}$, and $\alpha=0.08\,\text{fm}^4$. 
The dashed bars correspond to $N_{\text{max}}=8$ calculations.
Bands in the last but one column on the right indicate the cutoff dependence of the EGM potential. Experimental excitation energies are taken from~\cite{NNDC}. 
 }
\label{fig:itncsm_Li6_Spec}
\end{figure*}

We start with the simple nucleus \elem{Li}{6}. Figure~\ref{fig:itncsm_Li6_Spec} shows the excitation energies of the first four positive parity states obtained with the different chiral NN and NN+3N interactions. To indicate the uncertainty due to the convergence with respect to the model space we compare the results at $N_{\text{max}}=8$ (dashed bars) and $N_{\text{max}}=10$ (solid bars). 
The calculations are carried out at $\hbar\Omega=16\,\text{MeV}$ and the SRG evolution up to $\alpha=0.08\,\text{fm}^4$ is performed at the three-body level. 
As illustrated for the \elem{C}{12} spectrum in Ref.~\cite{MaVa14} once the SRG evolution is performed consistently at the three-body level, excitation energies show a negligible flow-parameter dependence. This remains true even for heavier systems, where the absolute energies show a sizable flow-parameter dependence~\cite{RoCa14,BiLa13b,RoLa11}. Since the SRG induced beyond-3N contributions predominantly cause an overall shift of all energies, their impact cancels out for the excitation energies.
In addition to the spectra for the individual interactions including the initial NN and NN+3N part, respectively, Fig.~\ref{fig:itncsm_Li6_Spec} also shows a combined spectrum for all the EGM interactions, where the bands indicate the spread of the excitation energies obtained with the different cutoffs. We observe that the excitation energies obtained with the EM and the $\opt$ Hamiltonians typically fall into the bands extracted for the EGM interactions. 
 
\begin{figure*}[t]
\hspace*{-20pt}\includegraphics[width=1.1\textwidth]{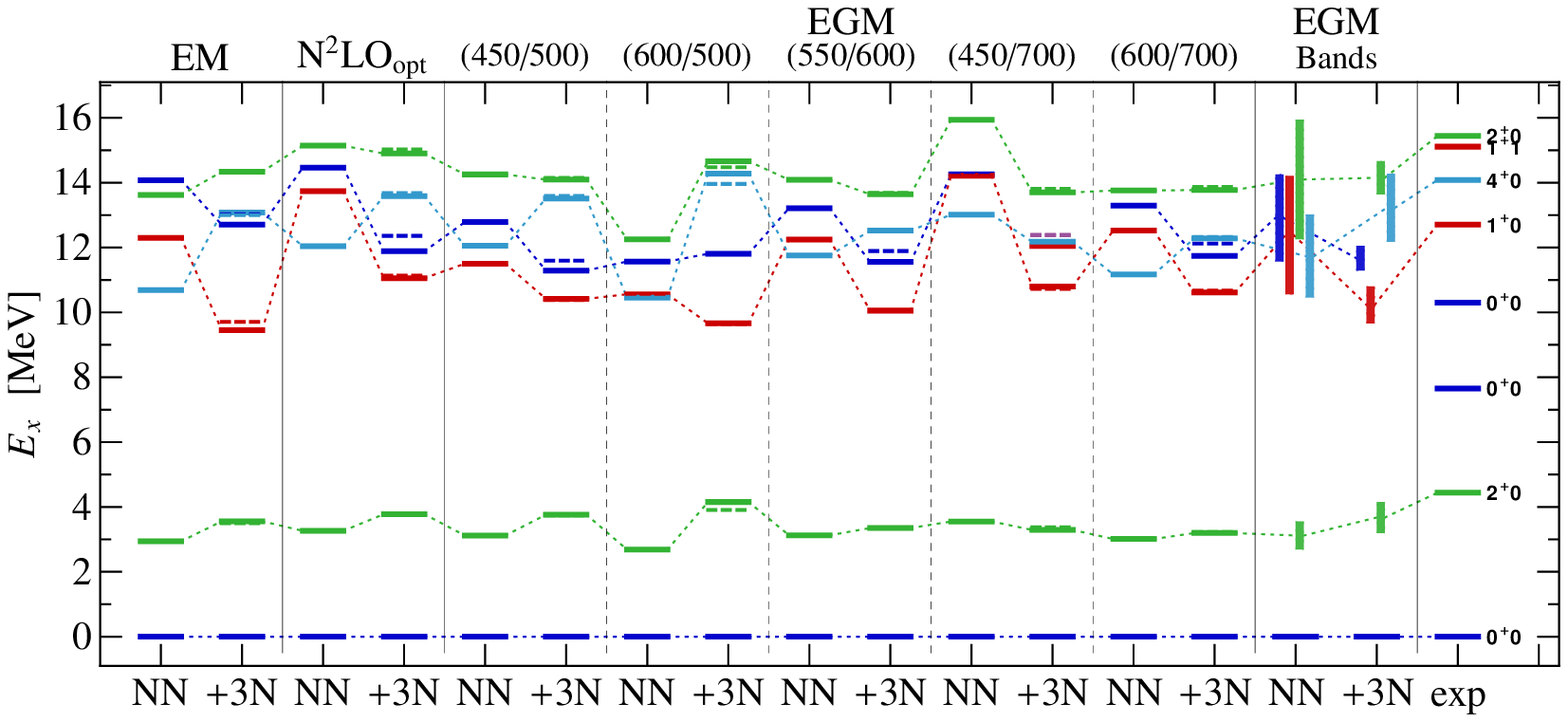}\\[-8pt] 
\caption{(color online) Excitation spectrum of  \elem{C}{12}. IT-NCSM calculations are performed for $N_{\text{max}}=8$ (solid bars) and $6$ (dashed bars), remaining parameters as in Fig.~\ref{fig:itncsm_Li6_Spec}. Experimental excitation energies are taken from~\cite{NNDC}.}
\label{fig:itncsm_C12_Spec}
\end{figure*}

A first inspection of the spectra reveals that the sensitivity of the different excited states to the Hamiltonian is quite different. Whereas the excitation energy of the first $0^+$ state is largely unaffected by the different choices of chiral interactions or the inclusion of the chiral 3N force, the excitation energies of the $3^+$ and the $1^+$ states show a sizable variation. The inclusion of the chiral 3N interaction causes a shift of the energies in the same direction for all Hamiltonians, leading to a higher $2^+$ and a lower $3^+$ excitation energy for the full NN+3N interactions. In the case of the EGM interactions, the band constructed from the cutoff dependence of the $3^+$ excitation energy nicely overlaps with the experimental energy.  For the $2^+$ excitation energy there is a clear discrepancy and the chiral 3N interaction shifts the state further away from the experimental energy in all cases. However, the first $2^+$ state in the experimental spectrum is a broad resonance and there is a narrow second $2^+$ state about $1$ MeV above. Thus the inclusion of continuum degrees of freedom, e.g., through the NCSM with continuum~\cite{NaQu16,LaNa15,BaNa13c}, will be important to understand and disentangle these $2^+$ states. The slow convergence of the calculated $2^+$ state might serve as a first indication for these continuum effects. For the $0^+$ excitation energy, the EGM band is closer to experiment although it does not overlap either---experimentally this state is a narrow resonance.

\begin{figure*}[t]
\hspace*{-20pt}\includegraphics[width=1.1\textwidth]{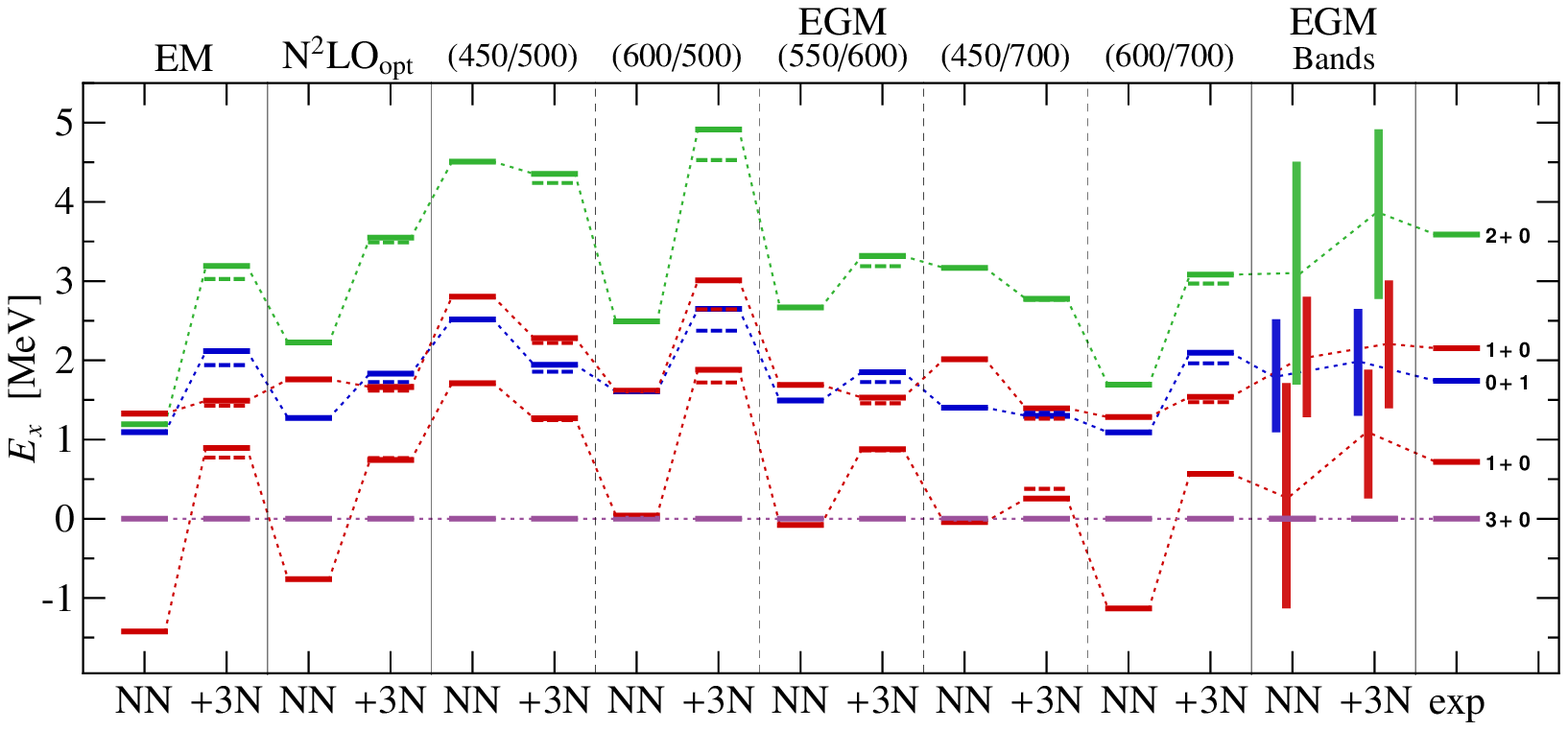}\\[-8pt]  
\caption{(color online) Excitation spectrum of  \elem{B}{10}. IT-NCSM calculations are performed for $N_{\text{max}}=8$ (solid bars) and $6$ (dashed bars), remaining parameters as in Fig.~\ref{fig:itncsm_Li6_Spec}. Experimental excitation energies are taken from~\cite{NNDC}.}
\label{fig:itncsm_B10_Spec}
\end{figure*}

Figure~\ref{fig:itncsm_C12_Spec} shows a similar analysis of the excitation spectrum of \elem{C}{12}. The excitation energies of the lowest positive-parity states obtained in IT-NCSM calculations are shown for all interactions. Obviously, the structure of excitation spectrum of \elem{C}{12} is richer than for \elem{Li}{6}. Previous investigations have shown that some of the excitation energies, e.g., for the first $1^+$ and $4^+$ states, are very sensitive to the 3N interaction. Furthermore, in comparison to experiment there are clear discrepancies of the $1^+$ excitation energy obtained for the EM interaction when including the 3N interaction~\cite{MaVa14,RoLa11}. The behavior of these states for different chiral interactions is, therefore, highly interesting. Note, the excited $0^{+}$ states are expected to have a distinct cluster structure that cannot be described accurately in tractable HO model spaces~\cite{Neff12}. Therefore, it is not clear whether the $0^+$ state obtained in the IT-NCSM corresponds the first excited $0^+$ state (Hoyle state) or the second one (see Ref.~\cite{MaVa14} for a more detailed discussion). For these reasons, we will not include this $0^+$ state into the following discussion on the sensitivity to the Hamiltonian. 
   
Comparing the spectra for the different interactions confirms the sensitivity of the $1^+$ and $4^+$ excitation energies to the underlying interaction, also the higher-lying $2^+$ state shows a large sensitivity. For all these states the sensitivity, as summarized by the bands for the EGM interactions, reduces significantly with the inclusion of the chiral 3N interaction. This might be interpreted as indication that the theoretical uncertainties are reduced when going from an incomplete chiral NN interaction to a complete and consistent chiral NN+3N Hamiltonian at N$^2$LO. For all interactions, the chiral 3N component shifts the $1^+$ states to lower excitation energies. As a result, all interactions underestimate the $1^+$ excitation energy by more than 2 MeV---even considering the uncertainty band, there is a clear discrepancy with experiment. Since all Hamiltonians employed here use local or non-local 3N interactions at N$^2$LO, it will be very interesting the see whether next-generation chiral 3N interactions at N$^3$LO can resolve this discrepancy.
In contrast to the $1^+$ and $4^+$ states, the energy of the first excited $2^+$ state shows very little sensitivity to the starting interaction and to the chiral 3N contribution. The bands extracted from the EGM interactions are small and the bands with and without the chiral 3N interaction overlap. Interestingly, all interactions tend to underestimate the $2^+$ excitation energy slightly.

It is also interesting to study the absolute \elem{C}{12} ground-state energies resulting from the different NN+3N interactions. The energies calculated with the EGM interactions span a range of $-96.8$ to $-80.5\,\text{MeV}$. This range contains the experimental energy of $-92.16\,\text{MeV}$.  Also the ground-state energy obtained with N$^2$LO$_{\text{opt}}$ interaction is within the EGM range, while the EM interactions predicts an energy of $-97.8\,\text{MeV}$ and thus the largest binding energy. However, it is important to note, that the energies are extrapolated from NCSM model spaces up to $N_{\text{max}}=8$ causing an estimated extrapolation uncertainty of about $1-2\,\text{MeV}$. What is more important is the impact of omitted SRG-induced 4N contributions that are sizable for the SRG flow-parameter $\alpha=0.08\,\text{fm}^4$ used here. From an analysis of the flow-parameter dependence we find that the 4N contributions are repulsive and, thus, will reduce the above binding energies. For instance, changing $\alpha$ for the EM interaction to $\alpha=0.04\,\text{fm}^4$, i.e., towards the bare interaction reduces the binding energy by about $2.3\,\text{MeV}$. Based on the flow-parameter dependence we cannot reliably estimate the binding energy expected for the bare interaction. Nevertheless, we can conclude that the absolute binding energies for the bare interactions will exhibit a spread of several MeV and tend to underestimate the experimental binding energies. This sensitivity of the absolute binding energies to the details of the interactions is consistent with the finding in~\cite{CaEk16} for the \elem{O}{16} ground-state energy.

As the final case, we discuss the excitation spectrum of \elem{B}{10} as shown in Fig.~\ref{fig:itncsm_B10_Spec}. The typical excitation energies for this odd-odd nucleus are much smaller than in \elem{C}{12}, therefore, shifts of the excitation energies of individual states by 1 MeV can change the spectrum drastically. Furthermore, full convergence of the excitation energies is more difficult to reach than for the \elem{C}{12} spectrum. Particularly the results with chiral 3N interactions show a residual $N_{\max}$ dependence, i.e., the excitation energies for $N_{\max}=8$ and $6$, indicated in Fig.~\ref{fig:itncsm_B10_Spec} by the solid and dashed levels, respectively, are slightly different. This residual $N_{\max}$ dependence is much smaller than the variations due to different Hamiltonians and, therefore, do not affect the present discussion.  

Already the first \emph{ab initio} calculations of \elem{B}{10} with 3N interactions have shown that the ordering of the first $3^+$ and $1^+$ states depends on the 3N interaction~\cite{CaNa02}. Many of the realistic NN interactions incorrectly predict the $1^+$ as ground state and only the 3N interaction restores the correct level ordering. This is also observed in Fig.~\ref{fig:itncsm_B10_Spec} for the chiral interactions---with one exception all chiral NN interactions predict the $1^+$ below or degenerate with the $3^+$ state. In all these cases, the chiral 3N interaction shifts the $1^+$ upwards relative to the $3^+$, thus, restoring the correct level ordering. An exception is the EGM interaction with cutoffs $(450/500)\,\text{MeV}/\text{c}$, which already gives the correct level ordering with the NN interaction, the chiral 3N interaction only leads to a slight reduction of all excitation energies. The EGM uncertainty band for the $1^+$ excitation energy is reduced by including the chiral 3N interaction and robustly indicates the $3^+$ as the ground state. Within the cutoff-uncertainty bands all excitation energies obtained with the chiral NN+3N are compatible with experiment. 

Our uncertainty analysis for the p-shell spectra provides a crucial verification of the predictive power of the chiral interactions. Besides distinct sensitivities of excitation energies to the 3N force, also systematic deviations from experiment beyond the theoretical uncertainty can be identified. These investigations identify the first $1^+$ state in \elem{C}{12} as an ideal benchmark for the next generation of chiral NN+3N interactions.

\section{Electromagnetic transitions and moments \label{sec:elmagObservables}}

Electromagnetic observables provide another window into structure of nuclei with different sensitivities and addressing complementary information. Therefore, we extend the discussion of sensitivities of nuclear observables to electromagnetic moments and transition strengths, focusing on electric quadrupole (E2) and magnetic dipole (M1) observables.

Coming from the discussion of excitation energies, several comments are in order: First, the convergence rate of electromagnetic observables, particularly of E2 observables, is significantly slower than the convergence of excitation energies. Owing to the sensitivity of the E2 operator on the long-range behavior of the wave function, large NCSM basis spaces are required to obtain the correct asymptotic behavior of the wave functions and to converge E2 observables. Eventually, one might still rely on extrapolations, e.g., within the novel schemes constructed in an effective field theory framework~\cite{OdPa15}, to extract a robust result.  

Second, the E2 or M1 operators should be transformed consistently with the Hamiltonian when using SRG transformations to improve the convergence behavior. The effect of this consistent SRG transformation of electromagnetic operators was only studied in a few selected cases. These studies indicated that the consistent SRG transformation changes electromagnetic observables only by a few percent~\cite{ScQu15,ScQu14}, which is why NCSM applications have not included these effect so far.   

Finally, chiral EFT also predicts the electromagnetic two-body current contributions consistently with the interactions. Also these contributions should be included for a complete treatment of electromagnetic observables. Pioneering calculation in a hybrid framework using chiral EFT currents with an Argonne interaction have indicated a significant influence of current contribution to electromagnetic moments and transition strengths~\cite{PaPi13}. 

Addressing all these effects in a comprehensive fashion will be the aim of our future studies of electromagnetic properties starting from consistent chiral EFT inputs. In a preparatory step towards the complete calculations, we study the impact of the Hamiltonian on E2 and M1 observables. As a novel aspect in the \emph{ab initio} context, we explore correlations between pairs of E2 or M1 observables involving the same states. As we will see in the following, the study of such correlations in an \emph{ab initio} framework can be extremely beneficial. Recently, correlations of E1 observables in closed-shell nuclei have been exploited for impressive predictions of observables sensitive to the charge and neutron distribution~\cite{HaEk16,MiBa16}.

\begin{figure}[t]
\hspace{-30pt}
\includegraphics[width=1.1\columnwidth]{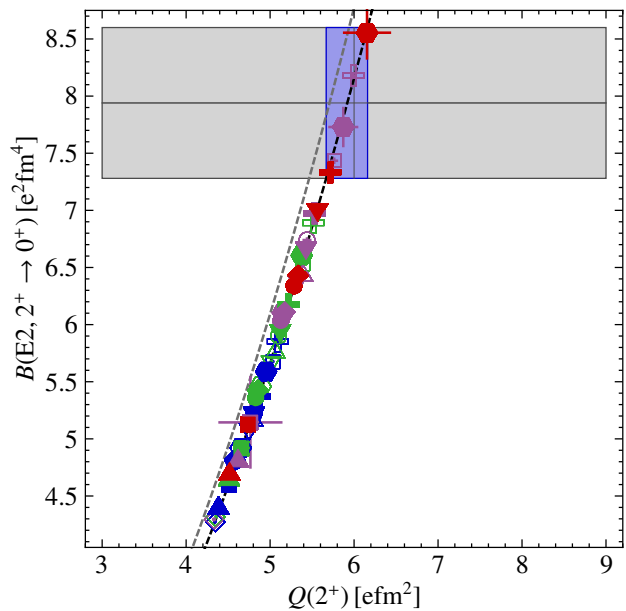}\\[-10pt] 
\vspace{10pt}
\caption{(color online) Correlation of quadrupole observables for the first $2^{+}$ state in \elem{C}{12}. Plotted is the reduced quadrupole transition strength $B(\text{E2},2^{+} \rightarrow 0^{+})$ to the ground state versus the quadrupole moment $Q(2^{+})$ obtained with different chiral NN (open symbols) and NN+3N interactions (solid symbols): EM (box), $\opt$ (circle), and EGM with cutoffs $(\LamChi / \LamSFR)=\{ (450/500),(600/500),$ $(550/600),(450/700),$ $(600/700) \}  \,\text{MeV}/\text{c}$ (diamond, triangle up, triangle down, hexagon, cross).
The IT-NCSM calculations are performed at $\hbar\Omega=16\,\text{MeV}$ and $\alpha=0.08\,\text{fm}^4$ using a model space of $N_{\text{max}}=2$ (blue), $4$ (green), $6$ (violet), and $8$ (red symbols). The error bars indicate the uncertainties of the threshold extrapolations in the IT-NCSM.
The dashed curves corresponds to the correlation obtained from formula~\eqref{eq:Qpole_BE2_corr} with a quotient of the intrinsic quadrupole moment set to one (grey) or fitted to theoretical data points (black).   
The grey shaded area indicates the error band of the experimental $B(\text{E2})$~\cite{RANe01}  and $Q$~\cite{VeEs83} value.
The blue shaded area corresponds to a prediction for $Q$ consistent with the theoretical correlation and the $B(\text{E2})$ measurement. 
 }
\label{fig:C12_BE2}
\end{figure}

We start with the discussion of E2 observables involving the first excited $2^+$ state and the $0^+$ ground state in \elem{C}{12}, i.e., the $B(\text{E2})$ transition strength form the $2^+$ state to the ground state and the quadrupole moment of the $2^+$ state. In Fig.~\ref{fig:C12_BE2} we present these two observables for the same set of chiral NN and NN+3N interactions used for the study of excitation spectra. In addition we show the results for different model-space truncations from $N_{\max}=2$ to $8$, which is important because of the slow convergence of these observables. Thus each symbol in the figure corresponds to a specific Hamiltonian at a specific value of $N_{\max}$. The grey rectangle indicates the experimental values for the $B(\text{E2})$ and the quadrupole moment including their experimental uncertainty. The uncertainty for the quadrupole moment is particularly large~\cite{VeEs83}, but new experiments are planned to reduce this uncertainty~\cite{PetriPrivComm}.

The picture that emerges from Fig.~\ref{fig:C12_BE2} is remarkable. All data points fall onto the same line, irrespective of the underlying chiral NN or NN+3N interactions and of $N_{\max}$. There is a strong and robust correlation between the two E2 observables emerging from our \emph{ab initio} calculations. The values of the individual observables show a sizable dependence on the underlying interaction and $N_{\max}$, but they always stay on the correlation line. As a general trend, with increasing $N_{\max}$ the quadrupole moment and the $B(\text{E2})$ continue to increase, indicating the slow convergence of these long-range observables. 

The robust correlation between this pair of quadrupole observables emerging from \emph{ab initio} calculations can be interpreted in terms of the simple rotational model by Bohr and Mottelson~\cite{BoMo75}, where both observables in the laboratory frame are connected to the intrinsic quadrupole moment $Q_{0}$ via the formulas  
\eq{ \label{eq:Qpole_BohrModelson}
  Q(J) = \frac{3K^{2} - J(J + 1)}{ (J + 1)(2J + 3)} Q_{0,s} \;,
}
and
\eq{ \label{eq:BE2_BohrModelson}
   B(\text{E2},J_{i} \rightarrow J_{f}) =\frac{5}{16\pi} Q^{2}_{0,t} \clebsch{J_{i}\,\,}{2\,}{J_{f}}{K}{0}{K} \;. 
}
Here $J$ is the angular momentum with the index $i$ and $f$ referring to the initial and final state, $K$ is the projection of the total angular momentum on the symmetry axis of the intrinsically deformed nucleus. 
For the investigated nuclei \elem{C}{12} and  \elem{Li}{6}, $K$ corresponds to the angular momentum of their ground states. 
The indices $s$ and $t$ of the intrinsic quadrupole moment indicate the "static" and "transition" observable $Q$ and $B(E2)$, respectively. One can combine both formulas such that the ratio of the intrinsic quadrupole moments ${Q_{0,t}}/{Q_{0,s}}$ is the only parameter that relates the two observables
\eqmulti{ \label{eq:Qpole_BE2_corr}
   B(\text{E2},J_{i} \rightarrow J_{f}) & = \frac{5}{16\pi}   \frac{\big{(} (J + 1)(2J + 3)\big{)}^{2}}{\big{(}3K^{2} - J(J + 1)\big{)}^{2}}  \clebsch{J_{i}\,\,}{2\,}{J_{f}}{K}{0}{K}  \\
  & \times \Big{(}\frac{Q_{0,t}}{Q_{0,s}}\Big{)}^{2} Q(J)^2 \;. 
}
In a rigid rotor model the intrinsic quadrupole moments $Q_{0,s}$ and $Q_{0,t}$ are expected to be equal. The correlation resulting from this assumption is represented by the grey dashed line in Fig.~\ref{fig:C12_BE2}, which slightly misses the correlation predicted in the \emph{ab initio} calculations. Using the ratio of the intrinsic quadrupole moments as a parameter to fit the above relations to the \emph{ab initio} results leads to ${Q_{0,t}}/{Q_{0,s}}=0.964$ and a correlation line that matches the \emph{ab initio} results perfectly, as seen from the black dashed line in Fig.~\ref{fig:C12_BE2}.

After having established this correlation in \emph{ab initio} calculations, we can exploit it to make predictions on one of the two observables based on experimental data for the other observable. In this particular case, the quadrupole moment of the $2^+$ state is poorly known, whereas the $B(\text{E2})$ has a much lower relative uncertainty. Thus we can use the experimental value and uncertainty $B(\text{E2})=7.94 \pm 0.66\,e^2\text{fm}^4$~\cite{RANe01} and translate it via the \emph{ab initio} correlation line into an value and uncertainty for the quadrupole moment of $Q(2^+)=(5.91 \pm 0.25)\,e\text{fm}^2$. The uncertainty of this value is one order of magnitude smaller than the uncertainty of the direct measurement. 

It is also much better than the theory uncertainty for a direct calculation of the quadrupole moment. For $N_{\max}=8$ the different chiral NN+3N interaction predict quadrupole moments in the range from 4.5 to 6.2 $e \text{fm}^2$ (red filled symbols in Fig.~\ref{fig:C12_BE2}) and these values still increase with increasing $N_{\max}$. So the sensitivity to the interaction and the slow convergence lead to a substantial theory uncertainty, which is eliminated through the use of the correlation together with one experimental observable. The quadrupole moment is also consistent, but more precise than direct predictions by nuclear lattice simulations of $Q(2^+)=(6 \pm 2)\,e\text{fm}^2$ obtained at LO~\cite{EpKr12}.

\begin{figure}[t]
\hspace{-30pt}
\includegraphics[width=1.1\columnwidth]{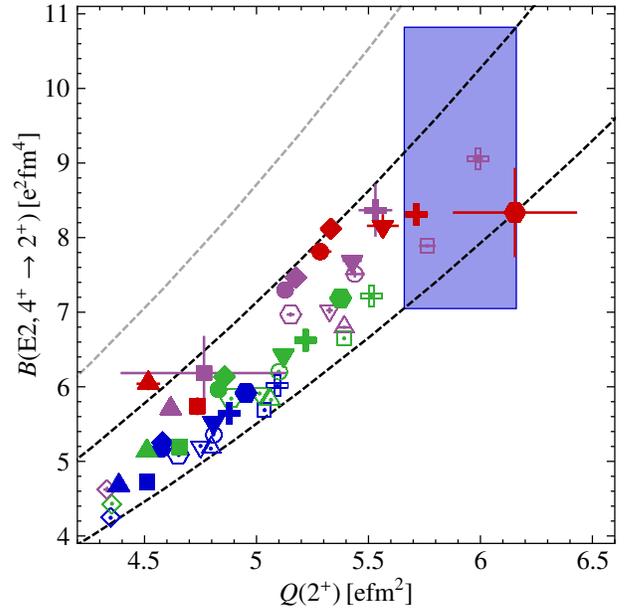}\\[-10pt] 
\vspace{10pt}
\caption{(color online) Correlation of the $B(\text{E2},4^{+} \rightarrow 2^{+})$ value and the quadrupole moment $Q(2^{+})$ in \elem{C}{12}. The parameters and definition of the symbols are as in Fig.~\ref{fig:C12_BE2}.
The black dashed lines mark the regime of the correlated theoretical data points. 
The blue shaded area corresponds to a prediction for the $B(\text{E2})$ transition strength consistent with the theoretical correlation and the predicted quadrupole moment from Fig.~\ref{fig:C12_BE2}.
}
\label{fig:C12_BE2_4plus}
\end{figure}

Due to the stability of the correlation in \elem{C}{12} one can also address higher excited states of the yrast band, as shown in Fig.~\ref{fig:C12_BE2_4plus}. In analogy to the previous correlation analysis we plot the $B(\text{E2},4^{+} \rightarrow 2^{+})$ as function of the quadrupole moment $Q(2^{+})$. The correlation motivated by the rotor model is present, but less clean. In particular, for the larger model spaces the theoretical data points start to spread around the fitted quadratic correlation curve. This indicates a more complicated structure of the $4^{+}$ state in large model spaces, deviating from the simple rotor model. Already the excitation energy of the $4^+$ state was much more sensitive to the interaction than the first $2^+$ state (cf. Fig.~\ref{fig:itncsm_C12_Spec}), indicating a different and more intricate structure for the $4^+$ state. 

Still, we can identify a correlation band indicated by the black curves covering almost all \emph{ab initio} results and parametrize it through the rotor model using a range for the ratio ${Q_{0,t}}/{Q_{0,s}}$ from $0.795$ to $0.905$, which differs significantly from the rigid rotor. Still, we can use this correlation band to predict the $B(\text{E2},4^{+} \rightarrow 2^{+})$ in the range from $7.05$ to $10.82\,e^2\text{fm}^4$, based on our previous extraction for $Q(2^+)$. This $B(\text{E2})$ is not known experimentally. Cluster models predict a value around $15\,e^2\text{fm}^4$~\cite{GaJe14} and a rigid rotor model with the intrinsic quadrupole deformation obtained from the 3$\alpha$ model applied in Ref.~\cite{AvGa07} gives $25\,e^2\text{fm}^4$~\cite{GaJe14}.

\begin{figure}[t]
\hspace{-30pt}
\includegraphics[width=1.1\columnwidth]{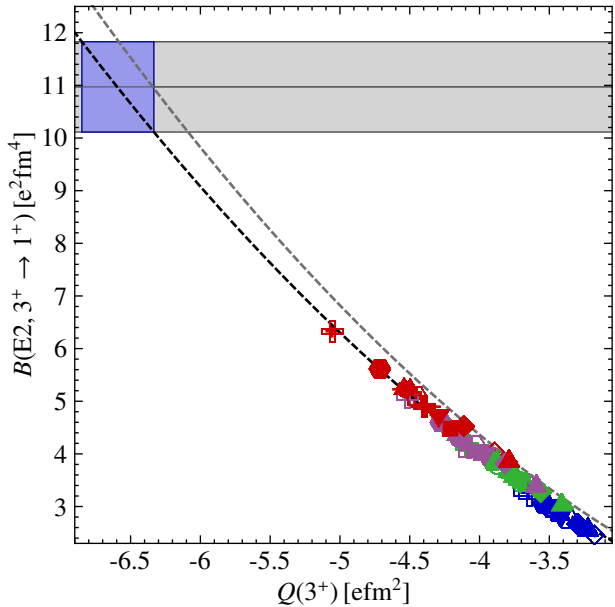}\\[-10pt] 
\vspace{10pt}
\caption{(color online) Correlation of quadrupole observables for the first $3^{+}$ state in \elem{Li}{6}. Plotted is the reduced quadrupole transition strength $B(\text{E2},3^{+} \rightarrow 1^{+})$ to the ground state as function of the quadrupole moment $Q(3^{+})$. The IT-NCSM calculations are  performed for $N_{\text{max}}=4$ (blue), $6$ (green), $8$ (violet), and $10$ (red symbols).
The remaining parameters and definition of the symbol shapes are as in Fig.~\ref{fig:C12_BE2}.
The grey shaded area indicates the error band of the experimental $B(\text{E2})$~\cite{TiCh02}. Note, there is no  measurement for the spectroscopic quadrupole moment.
} 
\label{fig:Li6_BE2}
\end{figure}
 
We now move to the lighter nucleus \elem{Li}{6} and repeat the correlation analysis. The lowest E2 transition is between the first excited $3^+$ state and the $1^+$ ground state. As we remarked earlier, the $3^+$ state is a narrow resonance and, therefore, the definition of the quadrupole moment is nontrivial. Since we are working in a bound-state approach, we can compute this observable nevertheless. In Fig.~\ref{fig:Li6_BE2} we show the correlation plot for the $B(\text{E2}, 3^{+} \rightarrow 1^{+})$ strength and the quadrupole moment of the first $3^{+}$ state in \elem{Li}{6}. Again we find a tight correlation between these two observables for all chiral NN and NN+3N interactions and model space truncations. A fit of the rotor model with ${Q_{0,t}}/{Q_{0,s}}=0.961$ again describes this correlation very well. 

Unlike the previous cases, all IT-NCSM calculations underestimate the $B(\text{E2})$ value---we only get about half of the experimental transition strengths. At the same time, there is a systematic dependence on the model-space truncation $N_{\max}$. For all interactions the absolute values of the quadrupole moment and the $B(\text{E2})$ increase monotonically with $N_{\max}$ with no indication of convergence. This general behavior is consistent with the findings in Ref.~\cite{NaVa01} using the CD-Bonn potential and hints at missing continuum effects for the description of the $3^+$ resonance. Still, we can use the rotor model to extract a value of the quadrupole moment of $Q=-6.59(26)\,\text{efm}^2$ based on the measured $B(\text{E2})$. 
        
\begin{figure}[t]
\hspace{-30pt}
\includegraphics[width=1.1\columnwidth]{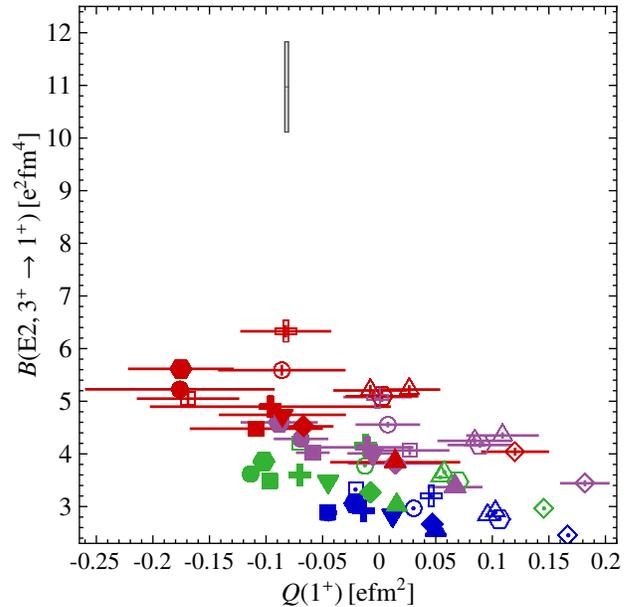}\\[-10pt] 
\vspace{10pt}
\caption{(color online) Correlation of quadrupole observables for the first $1^{+}$ state in \elem{Li}{6}. Plotted is the reduced quadrupole transition strength $B(\text{E2},3^{+} \rightarrow 1^{+})$ to the ground state as function of the quadrupole moment $Q(1^{+})$. The remaining parameters and definition of the symbol shapes are as in Fig.~\ref{fig:Li6_BE2}.
The grey shaded area indicates the error band of the experimental $B(\text{E2})$~\cite{TiCh02} and $Q(1^{+})$~\cite{CeOl98}.
} 
\label{fig:Li6_BE2_Q1}
\end{figure}

We can consider the same $B(\text{E2}, 3^{+} \rightarrow 1^{+})$ for \elem{Li}{6} in connection with the quadrupole moment of the $1^+$ ground state instead of the excited $3^+$ state. As shown in Fig.~\ref{fig:Li6_BE2_Q1} this pair of E2 observables does not exhibit a robust correlation. 
Because the measured quadrupole moment of $-0.08178(164)\,\text{efm}^2$~\cite{CeOl98} is so close to zero, the experimental data point in the correlation plot is incompatible with the rigid rotor model. The very small quadrupole moment of the ground-state is governed by more subtle structural effects rather than the robust effects from the rotor model. Note the impact of the importance truncation also becomes noticeable, because of the small magnitude of the quadrupole moment. Moreover, several theoretical quadrupole moments have a positive sign, i.e., predict a prolate deformation, but move with increasing $N_{\text{max}}$ towards the slightly oblate deformation as experimentally measured.
This example demonstrates, that the E2 correlations in \elem{C}{12} and \elem{Li}{6} that we identified from our \emph{ab initio} calculations and interpreted by the simple rotor model are non-trivial findings. 

\begin{figure}[t]
\hspace*{-23pt}
\includegraphics[width=1.1\columnwidth]{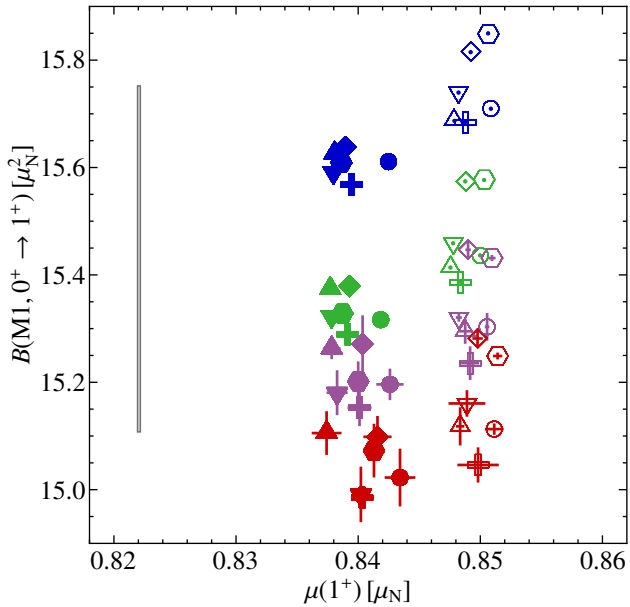}\\[-10pt] 
\vspace{10pt}
\caption{(color online) Correlation of magnetic-dipole observables for the $1^{+}$ ground state and the first $0^+$ excited state in \elem{Li}{6}. Plotted is the magnetic dipole transition strength $B(\text{M1},0^{+} \rightarrow 1^{+})$ to the ground state versus the magnetic-dipole moment $\mu(1^{+})$. The IT-NCSM calculations are  performed for $N_{\text{max}}=4$ (blue), $6$ (green), $8$ (violet), and $10$ (red symbols).
The remaining parameters and definition of the symbol shapes are as in Fig.~\ref{fig:C12_BE2}. The grey rectangle indicates the error band of the experimental $\mu(1^{+})$~\cite{BeBo74,Lutz67} and $B(\text{M1},0^{+} \rightarrow 1^{+})$~\cite{TiCh02} value.
}   
\label{fig:Li6_M1}
\end{figure}
  
Finally, we discuss an example for a different electromagnetic operator, the magnetic dipole or M1 operator. Unlike the E2 observables we discussed so far, the M1 operator does not depend on the spatial distance, but only probes the spin and orbital angular-momentum structure of the state. This leads to a different convergence behavior of M1 observables in NCSM-type calculations. Furthermore, from a macroscopic model build on an intrinsic state the magnetic dipole moment and the $B(\text{M1})$ transition strength show a less trivial but again quadratic relation depending on an effective and intrinsic $g$-factor~\cite{BoMo75}. Therefore, it is interesting to explore the spin and orbital structure that determines pairs of M1 observables in \emph{ab initio} calculations and to identify correlations.

In Fig.~\ref{fig:Li6_M1} we plot the $B(\text{M1},0^{+} \rightarrow 1^{+})$ transition strength from the excited $0^+$ state to the $1^+$ ground versus the magnetic dipole moment $\mu(1^{+})$ of the ground state in \elem{Li}{6}. As before, we consider the full set of interactions for a range of model space truncations $N_{\max}=4,6,8,$ and $10$. One should note that the range of $B(\text{M1})$ and $\mu$ presented in the plot, covering around 7\% relative change in both observables,  is very small compared to the typical variations of the E2 observables discussed before. This already indicates that M1 observables are more robust with respect to interaction and model-space choices. 

The picture regarding correlations is also different from the E2 observables, the calculations do not collapse on a universal correlation line. There are distinct groups of points with very systematic trends. First, the calculations using chiral NN+3N interactions (full symbols) are separated from calculations with only chiral NN forces (open symbols). The inclusion of the chiral 3N interaction reduces the dipole moment systematically by about 1\%, simultaneously the $B(\text{M1})$ is reduced by a similar amount. Second, with increasing $N_{\max}$ the $B(\text{M1})$ strength is systematically reduced, while the dipole moment remains practically constant. 
Third, the different input Hamiltonians for fixed $N_{\max}$ give very similar results as indicated by the groups of same-colored open or full symbols in Fig.~\ref{fig:Li6_M1}. In summary, both observables are robust with respect to the choice of chiral NN+3N interaction but they are influenced by the chiral 3N force. The dipole moment is converged while the $B(\text{M1})$ shows a systematic decrease with increasing $N_{\max}$ which might be related to the resonance nature of the $0^+$ state. 

Comparing the calculations to experiment, indicated by the narrow grey rectangle in Fig.~\ref{fig:Li6_M1}, provides an interesting perspective. The ground-state magnetic dipole moment of \elem{Li}{6} is known with an excellent accuracy from atomic physics measurements~\cite{BeBo74,Lutz67}. The calculations with chiral NN+3N interactions deviate from experiment by about 2\%---though this is a small deviation by our standards, it is very systematic. The experimental uncertainty on the $B(\text{M1})$ are larger and most of the calculations fall within the error bar of the experiment. However, the systematic $N_{\max}$ dependence of the calculation suggests that the converged $B(\text{M1})$ will be outside the experimental error bar for all Hamiltonians. 

This is clearly a case where precision studies, both in experiment and in \emph{ab initio} theory will be very valuable. As mentioned in the beginning of this section, our studies of electromagnetic observables are not fully complete yet. We have not included the consistent SRG evolution of the electromagnetic operators and we have not included consistent electromagnetic two-body currents from chiral EFT. Both corrections enter as additive two-body pieces in the electromagnetic operators and they will affect both observables in the pairs of E2 and M1 observables discussed here. In case of the E2 observables, we expect the correlation line to be largely unaffected by these corrections. However, they will play a role for the specific values of the observables, particularly at the precision level of the M1 observables discussed above. This aspect will be a focus of our future studies.

\section{Conclusions \label{sec:conclusion}}

We have presented \emph{ab initio} IT-NCSM calculations for the spectroscopy of p-shell nuclei using a large set of different chiral NN+3N interactions. In this way we addressed the sensitivity of excitation spectra and electromagnetic observables to the input interactions, which constitutes a significant and previously unexplored contribution to the theory uncertainties of state-of-the-art \emph{ab initio} calculations. The variation of the input interactions also provides yet unexplored insights into the details of nuclear structure. 

We discussed the sensitivities of individual excitation energies and compared the resulting theory uncertainties to experiment. This provides an important diagnostic for the chiral interactions, particularly in case of a systematic disagreement with experiment beyond the uncertainties obtained from the different interactions. An example is the excitation energy of the first $1^+$ state in \elem{C}{12}, where the sensitivity to the different chiral NN+3N interactions is much smaller than the deviation from experiment. This will be an important test case for next-generation chiral interactions.

For electromagnetic observables the variation of the underlying interaction and of the model-space truncation allowed us to identify robust correlations between pairs of E2 observables merging from \emph{ab initio} calculations, which can be interpreted in terms of a rigid rotor model. These correlations offer a new tool to extract accurate predictions for \emph{ab initio} calculations. By combining the theoretically predicted correlations of two observables with a single experimental datum for one observable we can extract a value for the second observable that is far more accurate and robust than any direct \emph{ab initio} result. An example is the quadrupole moment of the first excited $2^+$ state in \elem{C}{12}, that we predict to be  $Q(2^+)=(5.91 \pm 0.25)\,e\text{fm}^2$ based on the experimental value of the associated $B(\text{E2})$. 

This work is a preparatory step towards a full quantification of theory uncertainties based on consistent inputs from chiral EFT. For example, using the new family of chiral interactions from LO to N$^4$LO developed within the LENPIC collaboration~\cite{EpKr15,EpKr15b,BiCa15}, we will be able to study the order-by-order systematics of nuclear observables and thus propagate the EFT uncertainties consistently to the level of nuclear structure observables. Including two-body currents and consistent SRG-evolution for the electromagnetic observables will be another important milestone towards precision \emph{ab initio} calculations that exploit the full potential of chiral EFT.


\section*{Acknowledgments}

We thank Marina Petri and Alfredo Poves for helpful discussions.
This work is supported by the DFG through grant SFB 1245, the Helmholtz International Center for FAIR (HIC for FAIR), the BMBF through contracts 05P15RDFN1 (NuSTAR.DA) and 05P2015 (NuSTAR R\&D).
TRIUMF receives funding via a contribution through the Canadian National Research Council. 
Numerical calculations have been performed at the computing center of the TU Darmstadt (LICHTENBERG), at the J\"ulich Supercomputing Centre, at the LOEWE-CSC Frankfurt, and at the National Energy Research Scientific Computing Center supported by the Office of Science of the U.S.~Department of Energy under Contract No. DE-AC02-05CH11231.

%

\end{document}